\documentclass[article,12pt]{article}

\usepackage{amsmath,aas_macros}
\usepackage[utf8]{inputenc}
\usepackage{fullpage}
\usepackage{natbib}
\usepackage[normalem]{ulem} 
\usepackage{graphicx,multicol,subfig,longtable,xcolor,lmodern}

\usepackage{hyperref}
\usepackage{natbib}
\usepackage[margin=1in]{geometry}

\date{}

\title{Intermittent turbulent fluctuations in solar coronal mass ejections}
\author{Apurva Bhagat, Sumit Tambe, Debesh Bhattacharjee and Prasad Subramanian}

\begin{document}

\maketitle

\begin{abstract}
Localised regions of high intensity fluctuations are known to be signatures of intermittency in fluid and plasma turbulence. We investigate such turbulent spots using near-Earth {\em in-situ} spacecraft observations of a sample of 125 solar coronal mass ejections (CMEs). We present statistical results which suggest that the intensity of the strongest turbulent spot and the turbulent spot occurrence rate are reliable indicators of the onset of the leading part of the CME event. Our findings also suggest that turbulent spots can be sites of enhanced proton heating. The findings of this study can enhance our understanding of intermittence in collisionless plasma turbulence and can improve CME/sheath-driven space weather impact prediction models.
\end{abstract}


\section{Introduction}
Earth-directed solar coronal mass ejections (CMEs) are known to often cause near-Earth space weather disturbances, many of which have deleterious consequences on satellite and ground-based technologies such as single event upsets in satellite electronics, enhanced satellite drag, GPS disruptions and many more. Consequently, there is a large body of research that seeks to investigate the salient characteristics of CMEs that most influence the severity of space weather disturbances \citep{2021LRSP...18....4T,2022FrASS...917103B}. 
We focus here on one such characteristic in CMEs - the occurrence of intermittent, localized, broadband fluctuations, which we call turbulent spots---owing to their resemblance to turbulent spots commonly observed in fluid mechanics studies \citep{Turbulent_spots_AnnualRev}. 
While the solar wind in itself is well-known to be turbulent (e.g., \citealp{2013LRSP...10....2B,2016JPlPh..82f5302C} and references therein) and intermittency in solar wind and space plasma turbulence is well recognized (\citealp{2015JPlPh..81d3901Z,2019ApJ...884L..57P, 2025ApJ...985..113W}), there are only a few studies regarding intermittent turbulent fluctuations in CMEs of which we are aware (e.g., \citealp{2013AnGeo..31.1559K,2025ApJ...986L..27R, 2023marquez}). 

Intermittency in turbulent fluctuations of a quantity (e.g. the magnetic field or velocity) are localized events which typically occur in the high wavenumber tails of the fluctuation spectrum, representing deviations from Gaussianity \citep{2013LRSP...10....2B}.
There are other, related definitions of an intermittent structure in plasma turbulence (see \citealp{2023PhPl...30d0502V} for a review); for instance, \citep{2016PhPl...23e5705Z} consider them to be regions of enhanced current density, vorticity and energy dissipation.
Self-organization and intermittence in weakly collisional/collisionless plasmas (which is of interest to us) is thought to occur via the magneto-immutability effect (\citealp{2024JPlPh..90f5301M}) and result in particle energization (\citealp{2019JPlPh..85c1702M}).
Coherent velocity field structures are also observed in neutral fluid flows -- in wall bounded turbulent boundary layers (e.g., \citealp{2021JFM...911A...2B}) and in turbulent pipe flow (\citealp{2023JFM...971A...9G}), the interpretation of which derives from Townsend's wall-attached eddy hypothesis (\citealp{Townsend_1951}). In laminar-turbulent boundary layer transitions, velocity field intermittency is often quantified as the time-fraction occupied by the turbulent spots, and is known to follow the universal Dhawan-Narasimha intermittency curve \citep{Dhawan_Narasimha_1958}.

We aim to study the strength and characteristic occurrence frequencies of turbulent spots in different parts of Earth-directed CMEs, using {\em in-situ} data from 125 well observed events. $\S$ \ref{sec:In_situ_data} describes the specifics of the data, which is analysed using wavelet transforms to identify regions of extreme intermittent fluctuations associated with turbulent spots. $\S$ \ref{sec:wavelet} discusses the analysis method and results. Based on the wavelet analysis, we formulate statistical discriminators between different parts of the CME event, with special emphasis on discriminating between the solar wind background and the sheath. This is detailed in $\S$ \ref{sec:occurence of turbulent spots}. 
We summarize and discuss the results in $\S$ \ref{Sec: Summary_and_discussion}.

\section{In-situ data from CMEs observed near the Earth}
\label{sec:In_situ_data}
\begin{figure}
  \centering
  \begin{minipage}[b]{0.45\textwidth}
    \centering
    \includegraphics[width=\textwidth]{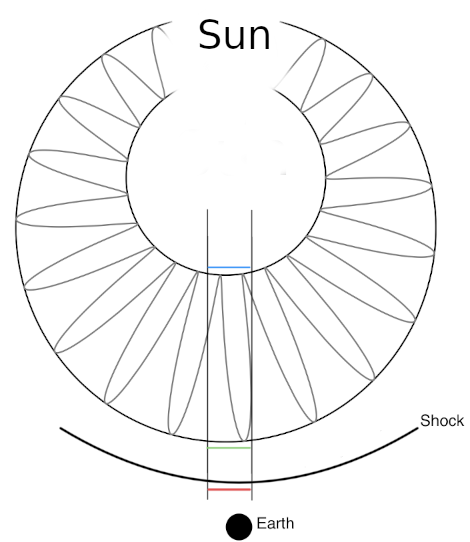}
  \end{minipage}
  \hfill
  \begin{minipage}[b]{0.45\textwidth}
    \centering
    \includegraphics[width=\textwidth]{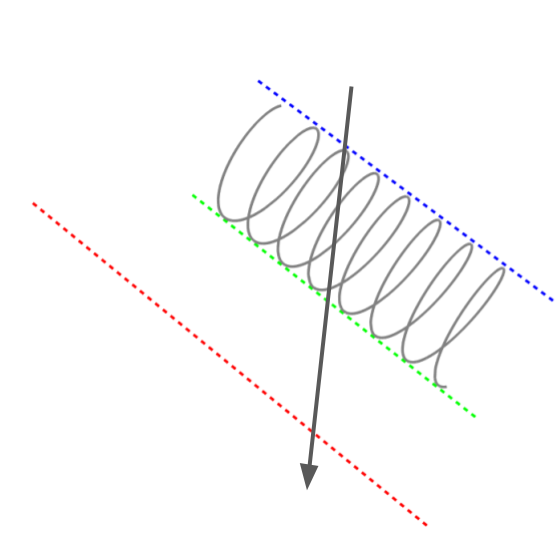}
  \end{minipage}
  \caption{Schematic of a flux rope CME. The left panel depicts an overall view of the flux rope CME. The torus-like lines denote the boundaries of the expanding CME and the grey lines denote the large-scale helical magnetic field that characterizes the flux rope. The CME typically drives a shock ahead of it, which is denoted by a circular arc. The dimensions of the Earth (while not shown to scale) are far smaller than the local dimensions of the flux rope CME. The vertical lines denote the limited extent of the structure that is encountered by near-Earth spacecraft which make in-situ measurements. The near-Earth, local extent of the shock is denoted by the red line and the front and rear boundaries of the CME are denoted by the green and blue lines respectively. The right panel is a zoomed-in view of the part of the structure between the vertical lines in the left panel. The right panel shows that the local view of the structure passing the Earth appears like a cylinder. The grey lines depict the helical magnetic field lines on the periphery. The red dashed line denotes the shock, the dashed green lines denotes the front boundary of the CME and the dashed blue line denotes the rear boundary of the CME. The positions of the shock, front and rear boundaries are shown using these linestyles in subsequent figures. The arrow denotes the direction of the CME velocity as it approaches the Earth. The near-Earth spacecraft (whose dimensions are negligible in comparison to that of the cylinder radius) sample the structure and make in-situ observations along this line. The arrow is intentionally titled to emphasize the fact that the spacecraft line of intercept need not coincide with the diameter of the cylinder.}
  \label{cartoonfig}
\end{figure}

An Earth-directed CME can be thought of as an organized magnetic structure riding on the background solar wind \citep{2011chen, 2019teresa}. CMEs propagate at speeds that usually exceed the local sound as well as Alfven speed. They therefore typically drive a shock ahead of them. The large-scale magnetic field in this structure is thought to resemble a helical flux rope 
\citep{2003chen,2018teresa_mc,2022debesh, 2001low, 1990burlaga} (left panel,fig~\ref{cartoonfig}). The local structure of a typical CME near Earth can be thought of as a cylinder (right panel,fig~\ref{cartoonfig}). 
The diameter of the flux rope near the Earth is typically $7$--$10,000$ times larger than the radius of the Earth. Spacecraft such as \href{https://wind.nasa.gov}{Wind}, which are situated at the L1 Earth-Sun Lagrange point, intercept the CME and it's associated shock, providing in-situ observations of quantities such as the plasma velocity, proton density, temperature and magnetic field along the line of intercept. The right panel of fig~\ref{cartoonfig} shows a possible line of spacecraft intercept, which might not always coincide with the diameter of the local flux rope cylinder.


\begin{figure}
\includegraphics[width=1\textwidth]{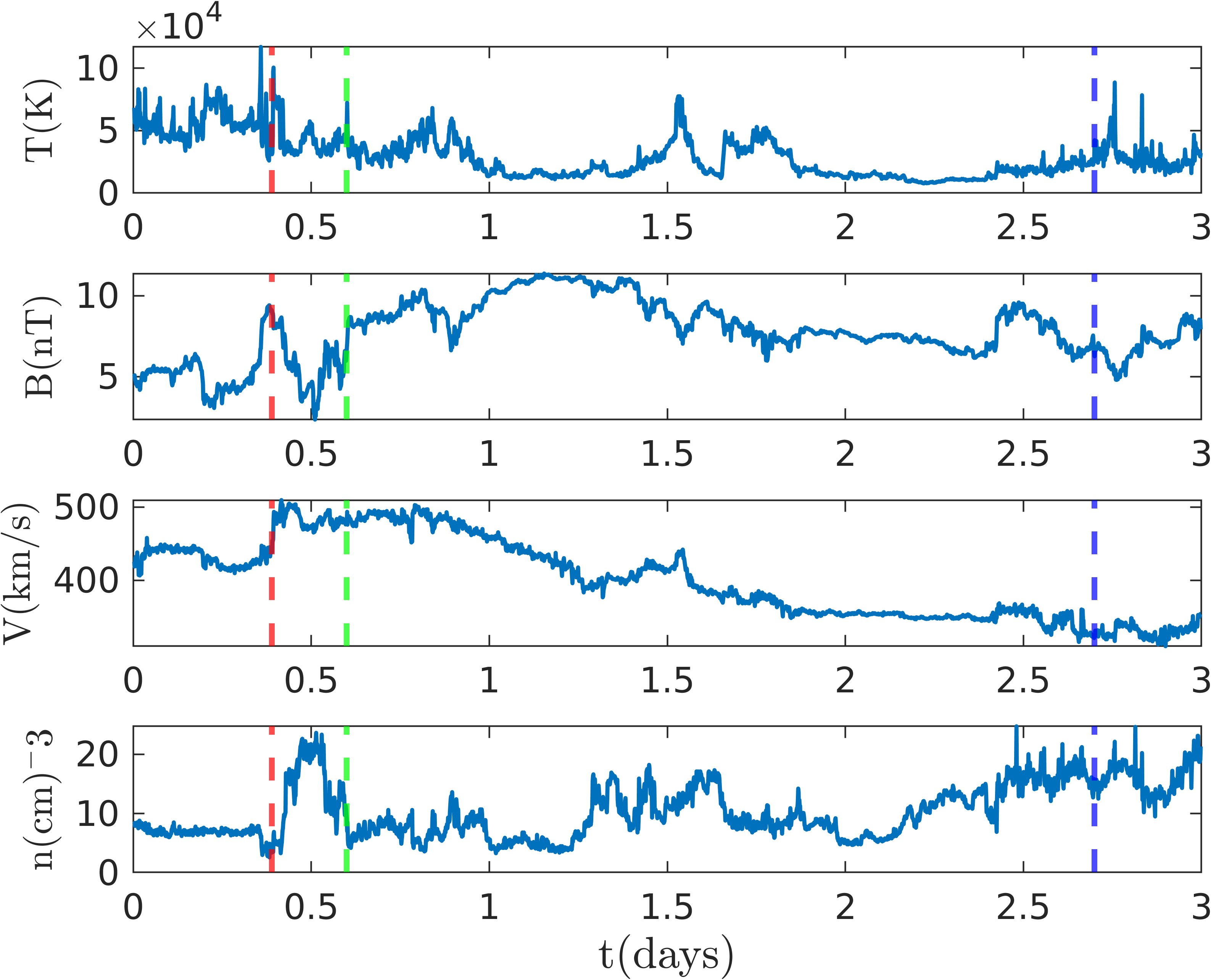}
\caption{Near-Earth time series measurements of proton temperature in $K$ (top panel), the magnetic field magnitude in $nT$ (second panel from top), the plasma velocity in ${\rm km/s}$ (third panel from top) and the proton number density in ${\rm cm^{-3}}$ (fourth panel from top) for the third event in dataset~\href{https://doi.org/10.5281/zenodo.18490716}{A1}, which occurred on 1995/06/30. The spacecraft intercepts the CME flux rope structure along an intercept line (shown by the arrow in panel (b) of figure~\ref{cartoonfig}), providing a one-dimensional cut along the structure. The red dashed line denotes the position of the shock, the green dashed line the front of the CME and the blue dashed line the rear of the CME. 
A one hour stretch immediately prior to the start of the event (in this case, the red dashed line) is taken to be the pre-event solar wind background.}
\label{timeseriesfig}
\end{figure}

The shock which is generally thought to be a fast mode shock \citep{2005manchester,2014kwon} is characterized by the (expected) jumps in physical quantities such as velocity, density and magnetic field. 
The magnetic flux rope structure of the CME that follows the shock is often modeled as a magnetohydrostatic cylinder \citep{1988burlaga, 2002hu}. Early models assumed a force-free cylindrical flux rope magnetic field structure (\citealp{1988JGR....93.7217B}) and subsequent ones relaxed the force-free assumption (e.g., \citealp{2019SoPh..294...89N}). In keeping with the expectations of such models, in-situ observations of CME interiors are characterized by an increase in the magnetic field, a smooth rotation of the magnetic field direction and a decrease in the proton plasma temperature \citep{1982klein}. We use such in-situ observations for 125 well-observed non-interacting Earth-directed CMEs, the details of which are listed in dataset~\href{https://doi.org/10.5281/zenodo.18490716}{A1}. This is a subset of the ICME list used in \citep{2023Debesh,2025debeshjaa}, where we have selected only those events that have a sheath. 
The start and end of the CME correspond to the boundaries of the modeled magnetohydrostatic structure \citep{2022debesh}. 
The region between the shock and the CME start is generally called the sheath \citep{2025debeshjaa}. 
Fig~\ref{timeseriesfig} shows a typical time sequence of in-situ observations with the shock, CME start and CME end marked. 
For concreteness, (as in \cite{2023debesh1,2025debeshjaa}) we define the ambient solar wind background as a 24 hour stretch immediately preceding the start of the event. 

Most of the attention in this field is devoted to the large-scale properties (as distinct from fluctuations) of quantities such as plasma velocity, proton temperature and magnetic field, although there are some which concentrate on turbulent fluctuations (\citealp{2020kilpua,2020good,2023Debesh, 2024shaikh,2025Deep,2025arXiv251204337E,2026arXiv260106904S}). 
We build on them, focusing on identifying intermittent turbulent spots in the sheath and interior of the CME, and comparing them to that in the background solar wind. 
As in \citep{2013AnGeo..31.1559K}, we use wavelet transforms to investigate the fluctuation characteristics of the magnetic field magnitude [$B(t)$, henceforth referred to as magnetic field], the magnitude of the plasma velocity [$V(t)$] and proton density [$n(t)$]. This is a well known technique that has been widely used in fluid dynamics to characterize turbulence and identify coherent structures in boundary layers (e.g., \citealp{2016PhPl...23e5705Z}). Following ideas used in the fluid dynamics literature, our focus here is on identifying coherent structures as a pathway to characterizing intermittency. 

\section{Turbulent spots}
\subsection{Identification}
\label{sec:wavelet}

\begin{figure}
\includegraphics[width=1\textwidth]{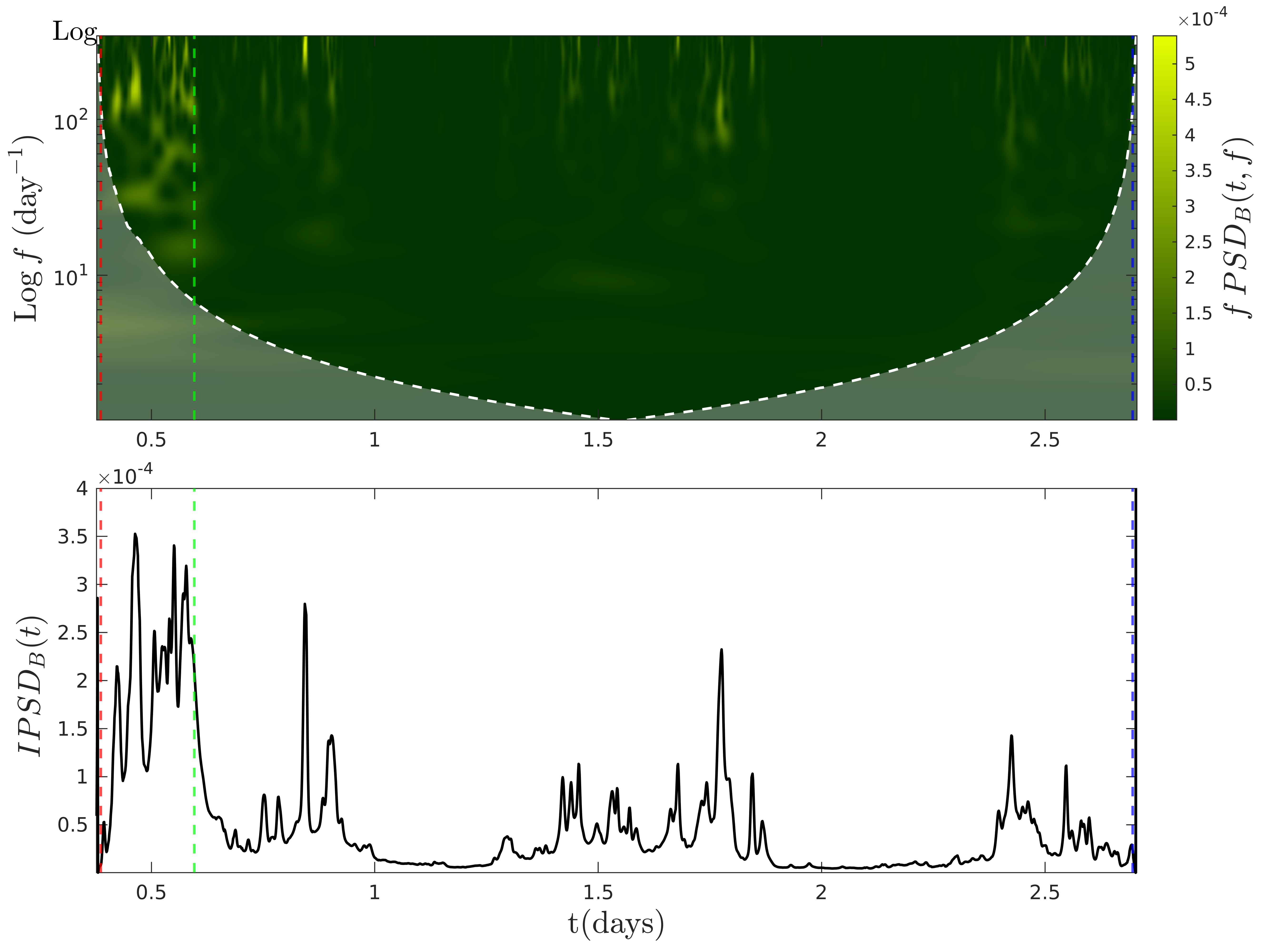}
\caption{The colorscale in top panel shows the frequency pre-multiplied power spectral density ($f\,PSD_{B}(t,f)$) for the Morlet wavelet transform of the magnetic field (second panel from top) for the observations shown in fig~\ref{timeseriesfig}. The x-axis shows time in days and the y-axis shows ${\rm log} f$ with $f$ in units of ${\rm day}^{-1}$. The power spectral density is defined in Eq~\ref{psdeq}, and $f \propto a^{-1}$. The white dashed line shows the cone of influence (coi). Data at frequencies below the coi are disregarded. The bottom panel shows the quantity $IPSD_{B}(t)$ (Eq~\ref{peakeq}). The red, green and blue dashed lines represent the sheath start, sheath end/CME start and CME end respectively, just as in figure~\ref{timeseriesfig}.}
\label{waveletfig}
\end{figure}

Recent studies in wall-bounded flows have highlighted the suitability of wavelet analysis in detecting intermittent fluctuations associated with turbulent spots \citep{Jiang_wavelet,Baars2015WaveletAnalysis,Simoni2016WaveletIntermittency,Anand2020TimeFrequency}. We use a continuous wavelet transform implemented in MATLAB \citep{Lilly2016jLab}. Following are some basic elements of the continuous wavelet transforms we use here. The wavelet transform of a signal $s(u)$ is defined as (\citealp{1998BAMS...79...61T,Cohen2003})
\begin{equation}
C(t,a) = \frac{1}{\sqrt{a}} \int s(u)\,\psi^{*} \, \biggl ( \frac{u - t}{a} \biggr ) \, du \, ,
\label{eq1}
\end{equation}
where the function $\psi$ is called the mother wavelet and $^*$ denotes the complex conjugate. 
The wavelet transform defined in Eq~\ref{eq1} is thus a convolution of the signal $s(u)$ with the conjugate of the mother wavelet function, with the quantity $a^{-1}$ playing the role of a frequency. 
In this work, we use the popular reduced form of the Morlet wavelet (\citealp{1998BAMS...79...61T}). 
\begin{equation}
\psi(t) = \pi^{-\frac{1}{4}}\,\exp [i \omega_{0} t]\,\exp[-t^{2}/2] \, ,
\label{cmoreq}
\end{equation}
where $\omega_{0}$ is a nondimensional frequency; $\omega_0$ is commonly taken to be equal to 6 to ensure the admissibility of the wavelet (i.e., the zero mean condition, down to the computer precision) \citep{annurev:/content/journals/10.1146/annurev.fl.24.010192.002143}. The wavelet shown in Eq~\ref{cmoreq} is thus a complex exponential modulated by a Gaussian (instead of a rectangular function as in a short-time Fourier transform \citep{Grochenig2001}). The one-sided wavelet power spectral density (PSD) at each timestamp $t$ as a function of the scale $a > 0$ (or equivalently frequency $ \propto a^{-1}$) for the wavelet spectrum is defined by 
\begin{equation}
PSD(t,a) \equiv \big | C(t,a) \big |^{2} \, .
\label{psdeq}
\end{equation}
We compute the PSD for $B(t)$, $V(t)$ and $n(t)$ for each event in our database. We denote the quantity for which the PSD is computed with an appropriate subscript; for instance, the PSD corresponding to the magnetic field [i.e., with $s(t) \equiv B(t)$ in Eqs~\ref{eq1}--\ref{psdeq}] would be $PSD_{B}(t,a)$. The quantities $PSD_{V}(t,a)$ and $PSD_{n}(t,a)$ can be defined analogously. In what follows, we use the frequency $f$ (which is $\propto a^{-1}$) instead of the scale $a$. In other words, we work with quantities such as $PSD_{B}(t,f)$, $PSD_{V}(t,f)$ and $PSD_{n}(t,f)$. 
Since $\int PSD\, df = \int f \,PSD\, d {\rm log} f$, the quantity $f\,PSD$ represents the wavelet power per unit interval in ${\rm log} f$ space, as is the case with the top panel of fig~\ref{waveletfig}. Such frequency pre-multiplied spectra ($f\,PSD$) are commonly used in fluid dynamics (e.g., \citep{1999PhFl...11..417K,2023JFM...971A...9G}).

For example, the upper panel of fig~\ref{waveletfig} shows the frequency pre-multiplied complex Morlet wavelet co-efficient power ($f\,PSD_{B}$) for the magnetic field time series shown in the second panel from top, fig~\ref{timeseriesfig}. 
The yellow features in the upper panel fig~\ref{waveletfig} represent intermittent features that have excess power over a range of frequencies; we call such features {\em turbulent spots}. Such turbulent spots are localized; their extents are typically $\approx 1/30$ of that of the CME. We have generated similar plots for $PSD_{B}$, $PSD_{V}$ and $PSD_{n}$ for each event. 

The nature of the wavelet transform (Eq~\ref{eq1}) is such that the time series data [$s(t)$] is convolved with the mother wavelet. Since the data is limited in time extent, the convolution operation leads to a situation where edge effects become important; in other words, the wavelet transform is contaminated by artificially padded data from beyond the minimum and maximum time limits of the data stream. 
This is commonly addressed by the use of a cone of influence (\citealp{1998BAMS...79...61T,brainsci13010021}), which is depicted by the white dashed line in the upper panel of fig~\ref{waveletfig}. The parts of fig~\ref{waveletfig} below the cone of influence (COI) are influenced by edge effects and therefore unreliable. Low(er) frequencies are relatively more susceptible. As discussed below, we will only use data for frequencies above the value of the COI boundary. 

An examination of fig~\ref{waveletfig} (for instance) shows that the turbulent spots (yellow streaks) have varying frequency extents. In this paper, we are primarily interested in the (time) location of these turbulent spots and knowing which region (background solar wind, sheath or CME) they belong to. A convenient way to do this is by computing the quantity
\begin{equation}
IPSD_{X}(t) = \int_{f_{\rm coi}}^{f_{\rm max}} PSD_{X}(t,f)\,df \, ,
\label{peakeq}
\end{equation}
for each timestamp, where $X$ can refer to $B$, $V$ or $n$. 
The lower limit of the integral in Eq~\ref{peakeq} ($f_{\rm coi}$) is the value of the COI at that time instance. In doing so, we ensure that the relatively unreliable data below the COI is not used. The upper limit $f_{\rm max}$ is simply the maximum frequency present in the data. For instance, the lower panel of fig~\ref{waveletfig} depicts the quantity $IPSD_B(t)$ for each timestamp.

Figure~\ref{IPSDfig} shows the proton temperature, $IPSD_{\rm B}$, $IPSD_{\rm V}$ and $IPSD_{\rm n}$ for the event shown in figure~\ref{timeseriesfig}. The turbulent spots correspond to peaks in $IPSD_{\rm B}$, $IPSD_{\rm V}$ and $IPSD_{\rm n}$. A visual inspection of fig~\ref{IPSDfig} reveals the following features: the proton temperature increases noticeably at the start of the event (which corresponds to the red dashed line). $IPSD_{\rm B}$ shows an upward trend in the sheath (the region between the red and green dashed lines). The peaks of $IPSD_{\rm B}$, $IPSD_{\rm V}$ and $IPSD_{\rm n}$ in the sheath are higher than those immediately preceding the sheath. We also note that a peak in $IPSD_{\rm B}$ need not correspond to one in $IPSD_{\rm V}$ or in $IPSD_{\rm n}$, and vice-versa. We next aggregate statistical results from these quantities for all the events in our sample.

\subsection{Turbulent spots: peak intensity and occurrence rate}
\label{sec:occurence of turbulent spots}
\begin{figure}
  \centering
  \begin{tabular}{cc}
    \includegraphics[width=0.45\textwidth]{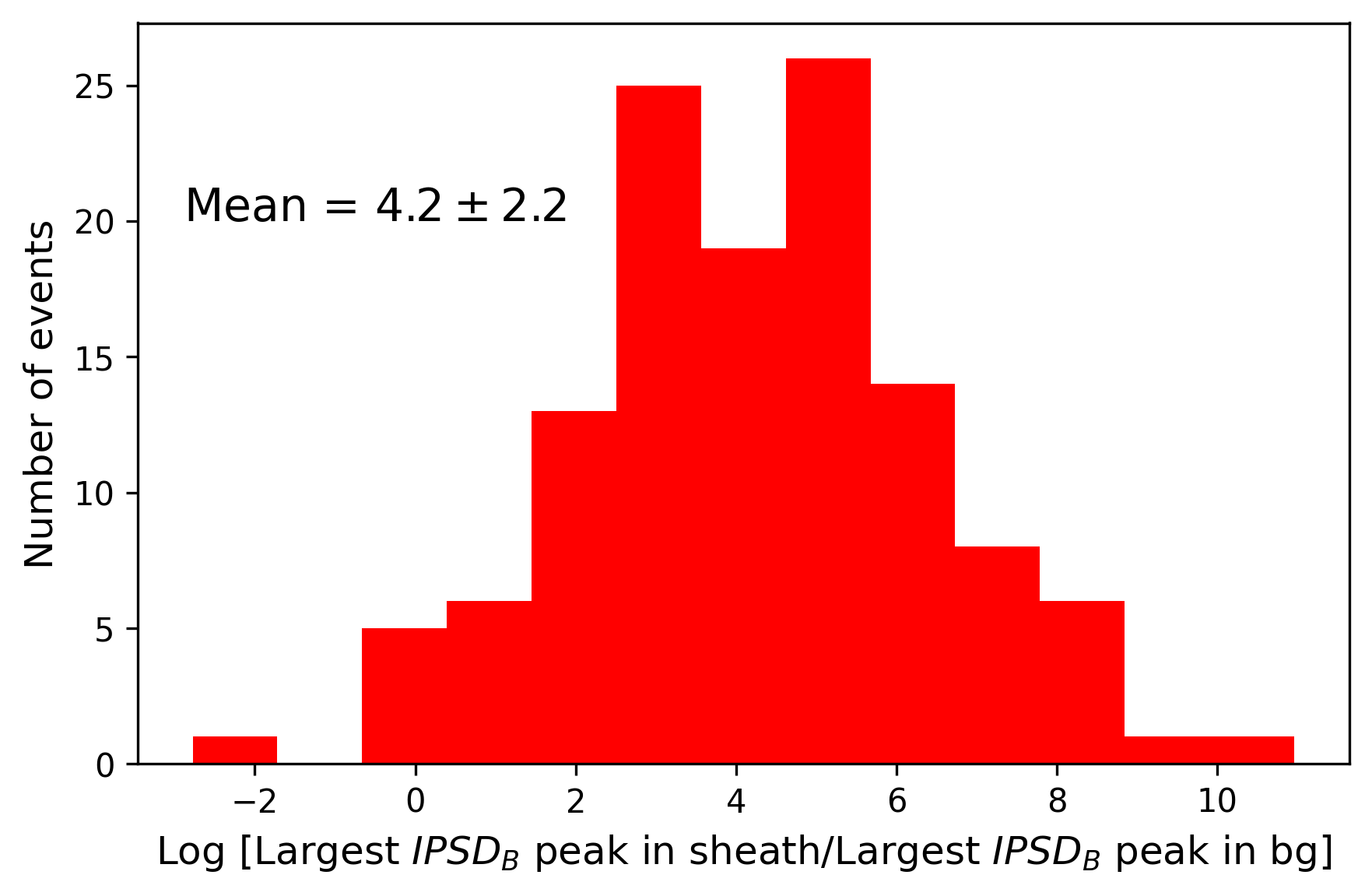} &
    \includegraphics[width=0.45\textwidth]{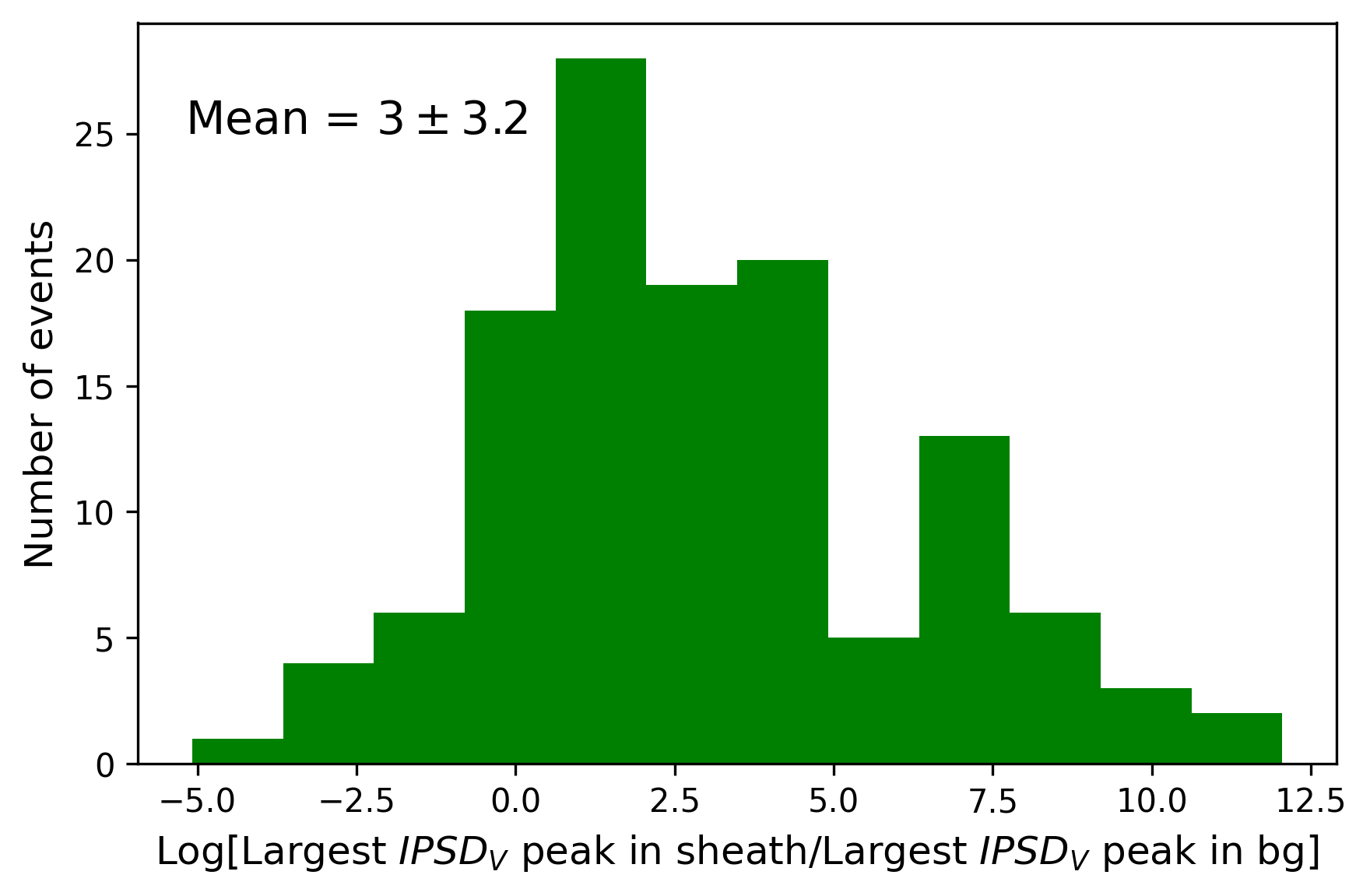} \\
    \includegraphics[width=0.45\textwidth]{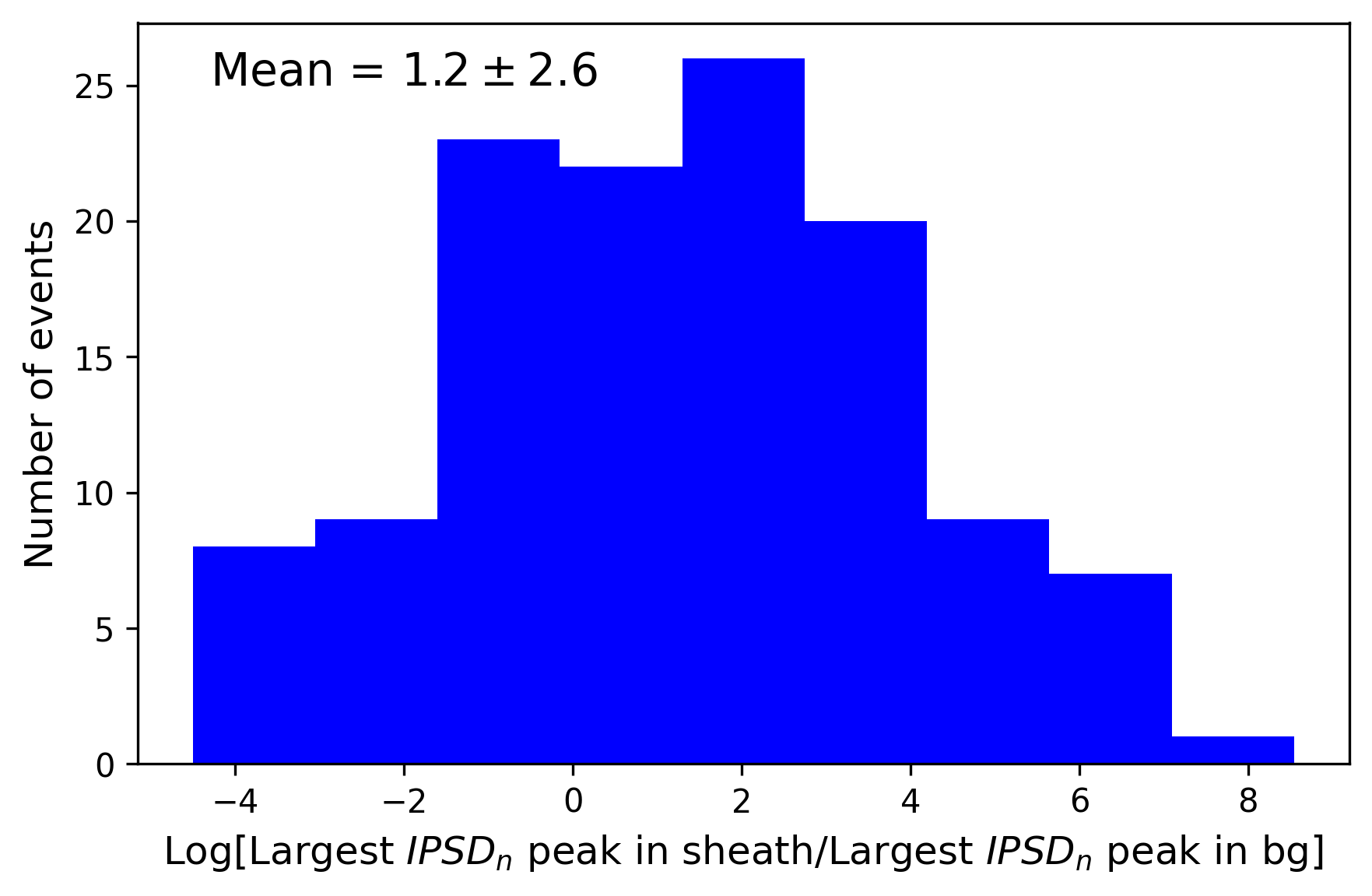} &
    \includegraphics[width=0.45\textwidth]{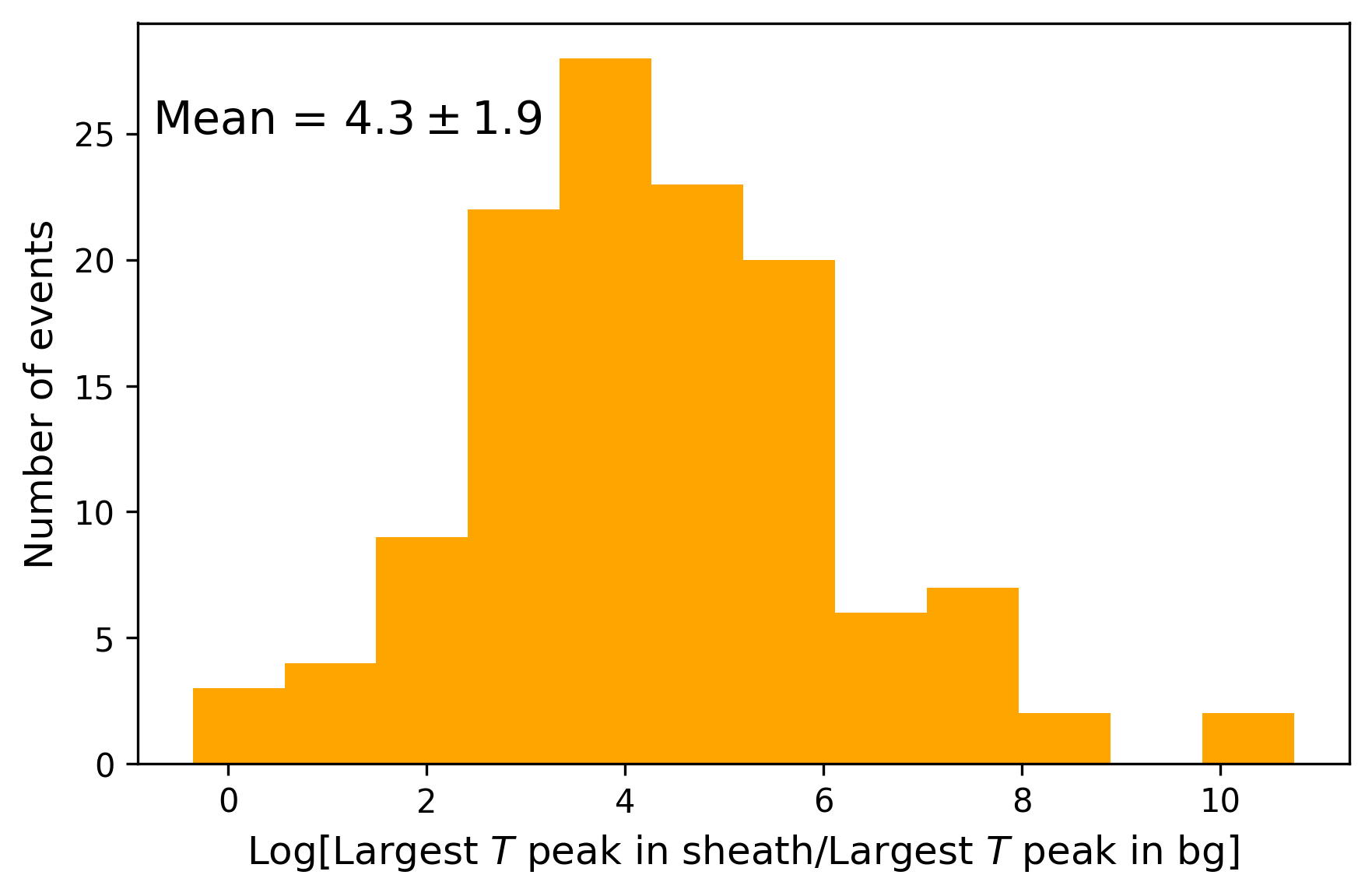}
  \end{tabular}
  \caption{Histograms for the logarithm of the ratio of the largest peak in the sheath to that in the solar wind background. The upper left panel (red) concerns $ISPD_{B}$, the upper right panel (green) concerns $IPSD_{V}$, the lower left panel (blue) concerns $IPSD_{n}$ and the lower right panel (yellow) concerns proton temperature ($T$).}
  \label{toppeakratios}
\end{figure}

To begin with, we concentrate on the highest peaks in each region for each quantity. For instance, we identify the highest peak in $IPSD_{\rm B}$ in the solar wind background, the highest peak in the sheath and the highest peak in the CME. In order to quantify the contrast between the background and the sheath, we compute the ratio of the highest $IPSD_{\rm B}$ peak in the sheath to that in the solar wind background. We repeat this procedure for the other quantities; i.e., $IPSD_{\rm V}$, $IPSD_{\rm n}$ and proton temperature ($T$). Histograms for the logarithm of these ratios for all the events in our sample are shown in figure~\ref{toppeakratios}. On the average, the logarithm of the ratio of largest peak of $IPSD_{\rm B}$ to that in the solar wind background is 4.2. The corresponding number for $IPSD_{\rm V}$ is 3, that for $IPSD_{\rm n}$ is 1.2 and that for the proton temperature is 4.3. Clearly, the intensity of the strongest turbulent spot (especially in $B$) is a reliable discriminator between the solar wind background and the sheath. The similarity in the histograms (both by way of general appearance and the mean value) for the temperature peak ratio (yellow, lower right panel) and the $IPSD_{\rm B}$ peak ratio (red, upper left panel) suggests that turbulent spots in $B$ might be sites of enhanced dissipation and consequent proton heating.

\begin{figure}
\includegraphics[width=1\textwidth]{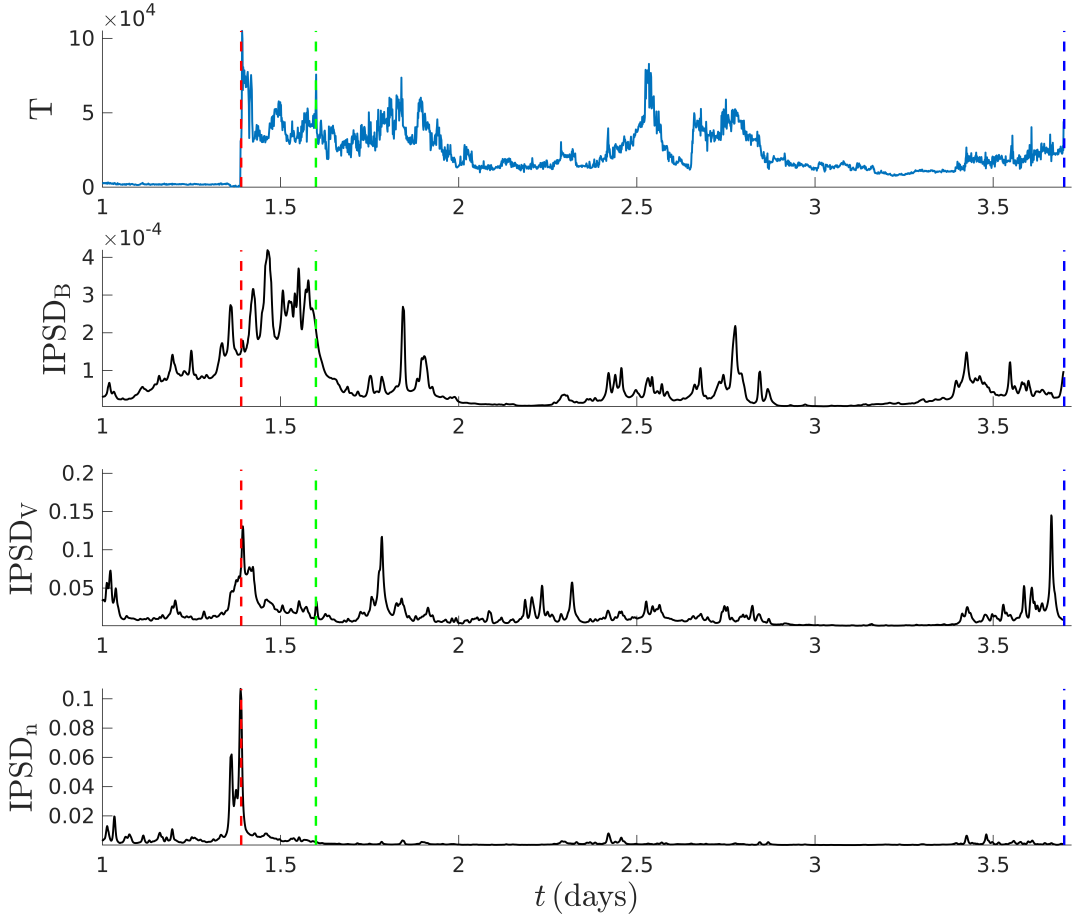}
\caption{The temperature (top panel), $IPSD_{\rm B}$ (second panel), $IPSD_{\rm V}$ (third panel) and $IPSD_{\rm n}$ (fourth panel) for the event shown in figure~\ref{timeseriesfig}.}
\label{IPSDfig}
\end{figure}


Figure~\ref{IPSDfig} shows that the proton temperature has noticeable peaks too, and some of the peaks are in the neighborhood of peaks in $IPSD_{\rm B}$, $IPSD_{\rm V}$ and $IPSD_{\rm n}$. This could be indicative of a connection between turbulent spots and dissipation/proton heating. In order to quantify this, we have computed the time interval between the largest proton temperature peak and the largest peak in $IPSD_{\rm B}$, $IPSD_{\rm V}$ and $IPSD_{\rm n}$. In the solar wind background, we find that the largest proton temperature peak occurs $\approx$ 12 minutes after the largest $IPSD_{\rm B}$ peak, 12 minutes after the largest $IPSD_{\rm V}$ peak and 10 minutes after the largest $IPSD_{\rm n}$ peak. In the sheath, the largest proton temperature peak occurs $\approx$ 13 minutes after the largest $IPSD_{\rm B}$ peak, 60 minutes after the largest $IPSD_{\rm V}$ peak and 43 minutes before the largest $IPSD_{\rm n}$ peak. In the CME, the largest proton temperature peak occurs $\approx$ 100 minutes after the largest $IPSD_{\rm B}$ peak, 10 minutes after the largest $IPSD_{\rm V}$ peak and 86 minutes after the largest $IPSD_{\rm n}$ peak. Taken together, these results provide further support for a scenario where the turbulent spots could be sites of enhanced dissipation and consequent proton heating, which is manifested as a proton temperature peak after a typical conduction timescale. Needless to say, it needs to be reconciled with details of dissipation processes and thermal conduction before arriving at a definite conclusion.

We next investigate the occurrence frequency of turbulent spots in each region. Since turbulent spots often seem to occur in clusters, we don't restrict our attention solely to the largest peak, as we did earlier. We separately consider the i) 5 ii) 10 and iii) 15 largest peaks in $IPSD_{\rm B}$, $IPSD_{\rm V}$ or $IPSD_{\rm n}$ in each region. If, for instance, we were considering only the 10 largest peaks, of which there were $N_{{\rm B\,bg}}$ peaks of $IPSD_{\rm B}$ in the background, $N_{{\rm B\,sheath}}$ peaks of $IPSD_{\rm B}$ in the sheath, $N_{{\rm B\,CME}}$ peaks of $IPSD_{\rm B}$ in the CME, with $N_{{\rm B\,bg}} + N_{{\rm B\,sheath}} + N_{{\rm B\,CME}} = 10$. We seek answers to questions such as: do turbulent spots occur more frequently in the sheath as compared to the background? To this end, we compute
{\begin{eqnarray}
\nonumber
PD_{{\rm B\,sheath}} \equiv \frac{N_{{\rm B\,bg}}}{T_{\rm bg}}\, , \,\, PD_{{\rm V\,bg}} \equiv \frac{N_{{\rm V\,bg}}}{T_{\rm bg}}\, , PD_{{\rm n\,bg}} \equiv \frac{N_{{\rm n\,bg}}}{T_{\rm bg}}\, , \\
\nonumber
PD_{{\rm B\,sheath}} \equiv \frac{N_{{\rm B\,sheath}}}{T_{\rm sheath}}\, , \,\, PD_{{\rm V\,sheath}} \equiv \frac{N_{{\rm V\,sheath}}}{T_{\rm sheath}}\, , PD_{{\rm n\,sheath}} \equiv \frac{N_{{\rm n\,sheath}}}{T_{\rm sheath}}\, , \\
PD_{{\rm B\,CME}} \equiv \frac{N_{{\rm B\,CME}}}{T_{\rm CME}}\, , \,\, PD_{{\rm V\,CME}} \equiv \frac{N_{{\rm V\,CME}}}{T_{\rm CME}}\, , PD_{{\rm n\,CME}} \equiv \frac{N_{{\rm n\,CME}}}{T_{\rm CME}}\, .
\label{peakdensityeq}
\end{eqnarray}}

\begin{figure}
\begin{center}

\includegraphics[width=0.6\textwidth]{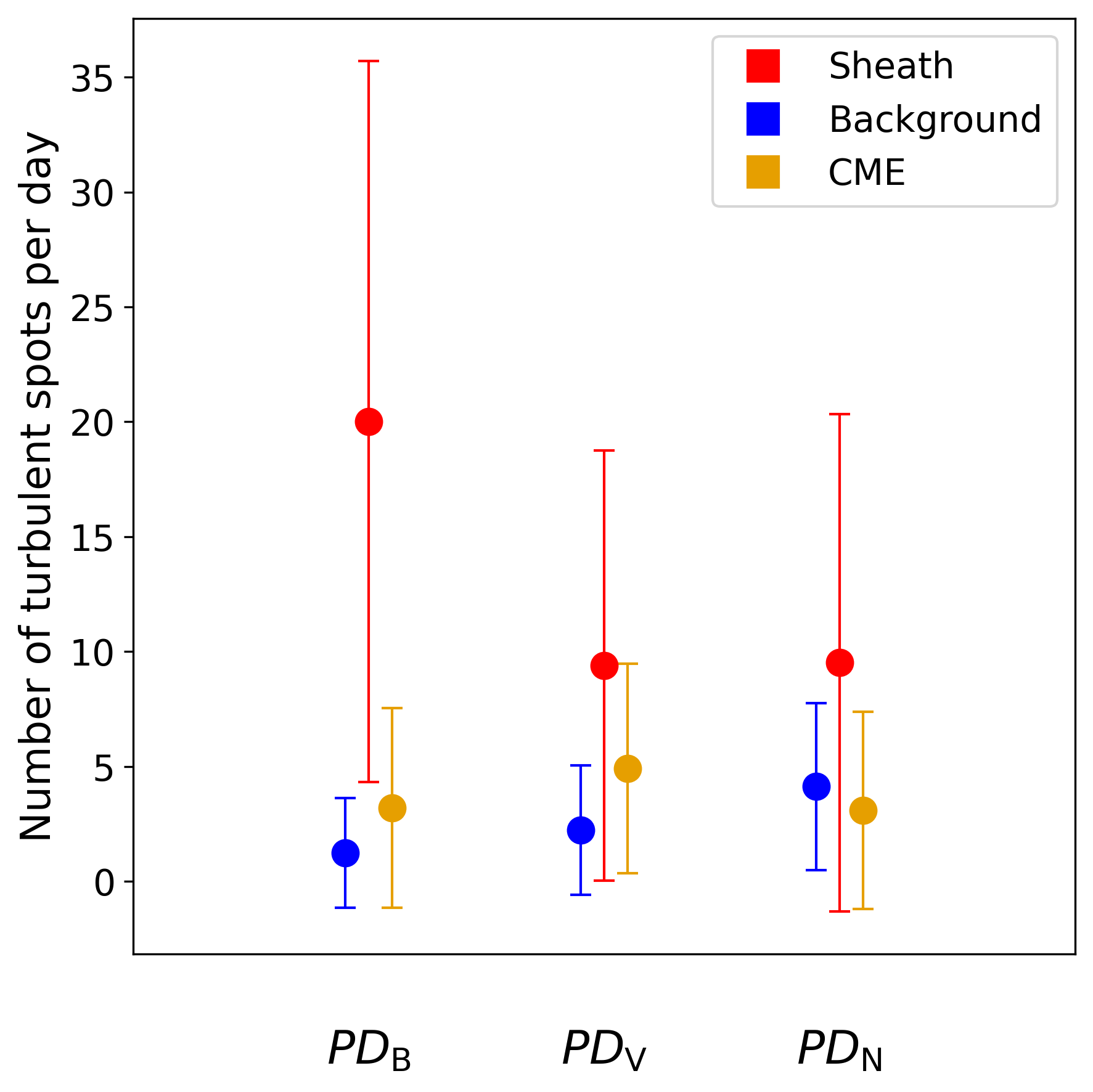}
\caption{The mean and standard deviation corresponding to the statistics for the 10 largest peaks shown in table~\ref{table-1}. For instance, the red dot and error bar denotes the mean of $PD_{B\,sheath}$ (which is 20.03 ${\rm day}^{-1}$) and its standard deviation (which is 15.69 ${\rm day}^{-1}$). The rest of the numbers are depicted similarly.}
\label{peakdensityhist}
\end{center}
\end{figure}


We compute the quantities listed in Eq~\ref{peakdensityeq} for each of the 125 events we have studied. Their means and standard deviations for the top 5, 10 and 15 peaks are listed separately in table~\ref{table-1}.

\begin{table}
\begin{center}

\begin{tabular}{lcccc}

{\textbf{Quantity}} & \textbf{No. of top peaks} & \textbf{Background}
 & \textbf{Sheath}
 & \textbf{CME }\\[12pt]
\hline
$P_{\rm D\,B}$
&  & $\langle P_{\rm D\,B\,bg} \rangle$
 &  $\langle P_{\rm D\,B\,sheath} \rangle$
 &  $\langle P_{\rm D\,B\,CME} \rangle $ \\[5pt]
& 5  & $  0.59 \pm 1.4\,{\rm day^{-1}}$
 & $  10.25 \pm 10.05 \,{\rm day^{-1}}$
 & $  1.58 \pm 2.2 \,{\rm day^{-1}}$ \\[5pt]
& 10 & $  1.25 \pm 2.39\,{\rm day^{-1}}$
 & $  20.03 \pm 15.69 \,{\rm day^{-1}}$
 & $  3.2 \pm 4.34 \,{\rm day^{-1}}$ \\[5pt]
& 15  & $  1.91 \pm 3.57\,{\rm day^{-1}}$
 & $  24.26 \pm 15.55 \,{\rm day^{-1}}$
 & $  3.55 \pm 5.3 \,{\rm day^{-1}}$ \\[12pt]
\hline
$P_{\rm D\,V}$
&  &  $\langle P_{\rm D\,V\,bg} \rangle$
 &  $\langle P_{\rm D\,V\,sheath} \rangle$
 &  $\langle P_{\rm D\,V\,CME} \rangle$ \\[5pt]
&5 & $ 1.16 \pm 1.8 \,{\rm day^{-1}}$
 & $ 4.67 \pm 6.23 \,{\rm day^{-1}}$
 & $ 2.55 \pm 2.67 \,{\rm day^{-1}}$ \\[5pt]
& 10  & $ 2.22 \pm 2.82 \,{\rm day^{-1}}$
 & $ 9.38 \pm 9.36 \,{\rm day^{-1}}$
 & $ 4.92 \pm 4.56 \,{\rm day^{-1}}$ \\[5pt]
& 15 & $ 3.09 \pm 4.84 \,{\rm day^{-1}}$
 & $ 13.13 \pm 12.45 \,{\rm day^{-1}}$
 & $ 4.24 \pm 4.25 \,{\rm day^{-1}}$ \\[12pt] 
\hline
$P_{\rm D\,n}$
&  & $\langle P_{\rm D\,n\,bg} \rangle$
 &  $\langle P_{\rm D\,n\,sheath} \rangle$
 &  $\langle P_{\rm D\,n\,CME} \rangle$ \\[5pt]
&5 & $ 2.06 \pm 2.06 \,{\rm day^{-1}}$
 & $ 4.83 \pm 6.27 \,{\rm day^{-1}}$
 & $ 1.56 \pm 2.08 \,{\rm day^{-1}}$ \\[5pt]
& 10 & $ 4.12 \pm 3.63 \,{\rm day^{-1}}$
 & $ 9.52 \pm 10.82 \,{\rm day^{-1}}$
 & $ 3.08 \pm 4.29 \,{\rm day^{-1}}$ \\ [5pt]
& 15 & $ 5.39 \pm 6.63 \,{\rm day^{-1}}$
 & $ 8.41 \pm 11.29 \,{\rm day^{-1}}$
 & $ 1.6 \pm 2.92 \,{\rm day^{-1}}$ \\[12pt]

\end{tabular}

\caption{Means and standard deviations of the quantities defined in Eq~\ref{peakdensityeq} for all the events in our sample for top 5, 10, and 15 peaks. }
\label{table-1}
\end{center}
\end{table}

Fig~\ref{peakdensityhist} is a graphical depiction of the numbers shown in table~\ref{table-1}. For clarity, we only depict the numbers corresponding to the 10 largest peaks. The figure may be understood as follows: for instance, (for the 10 largest peaks) the mean value of $PD_{{\rm B\,sheath}}$ is 20.03 turbulent spots per day and the standard deviation is 15.69 spots per day. This is represented by the red dot and error bar appearing above $PD_{{\rm B}}$. The other numbers are depicted similarly. 
It is evident from table~\ref{table-1} and fig~\ref{peakdensityhist} that the sheath has the most turbulent spots per unit time. It may be noted, however, the standard deviation for each of the quantities shown in fig~\ref{peakdensityhist} is comparable to the mean.

\section{Summary and discussion}

\label{Sec: Summary_and_discussion}
\subsection{Summary of results}
It is generally known that CME sheaths (and interiors, to a lesser extent) are more turbulent that the ambient background solar wind. Here we show that the enhancement is not uniform - rather, there are spatially localized regions of intense, wideband fluctuations which we call turbulent ``spots''. Such spots are a signature of intermittency, which is well known in the plasma physics and fluid dynamics communities. Using statistics from a sample of 125 well observed events, we offer novel ways of characterizing intermittence. 
Our main results are summarized in figure~\ref{toppeakratios}, table~\ref{table-1} and figure~\ref{peakdensityhist}. We find that

\begin{enumerate}
\item
Turbulent spots are localized (their sizes are typically 1/30 to 1/50 times that of the CME) regions of intense, broadband fluctuations. They correspond to peaks in curves such as that shown in the lower panel of fig~\ref{waveletfig}.
\item
There isn't a one-to-one correspondence between turbulent spots in $B$, $V$ and $n$. In other words, a turbulent spot in $B$ need not always be accompanied by one in $V$ or $n$ and vice-versa (fig~\ref{IPSDfig}).
\item
The intensity of the strongest turbulent spot in the sheath is considerably larger than the corresponding quantity in the solar wind background (figure~\ref{toppeakratios}), providing a clear and reliable discriminant between these two regions. Turbulent spots in $B$ provide the largest contrast in this regard. This result can potentially be used in an automated method to signal the onset of the sheath.
\item
The rate of occurrence of turbulent spots is substantially higher in the sheath as compared to the ambient solar wind. Turbulent spots in $B$ show the largest contrast (fig~\ref{peakdensityhist}) in this case too.
\item 
Turbulent spots, especially those in $B$, are associated with enhanced dissipation and proton heating.
\end{enumerate}

\subsection{Discussion}
We next address areas for further research arising from the work presented here. i) Our results can potentially serve as a useful input to automated methods for advance warning systems for space weather disturbances. Current space weather prediction models typically only use bulk parameters such as CME speed, the southward component of $B_{z}$ and others. Our work shows that the intensity of the strongest turbulent spot and the occurrence rate of turbulent spots can be reliable indicators of the onset of the sheath, which often triggers geomagnetic storms. This could augment the prediction capabilities of space weather prediction models, especially if implemented as on-board algorithms on future spacecraft positioned upstream of the Sun-Earth L1 point. ii) As a related matter, it is known that turbulence mediates reconnection processes that influence the coupling between CMEs and the Earth’s magnetosphere (\citealp{2003JGRA..108.1246B}). Our results on turbulent spots could serve as inputs for further understanding turbulence-mediated reconnection and their bearing on CME-magnetosphere coupling. iii) Finally, we note the similarities between the turbulent spots we have studied here and coherent structures identified in wall bounded fluid turbulence. On the face of it, CMEs are clearly not solid structures. However, there are studies that show that the excess magnetic pressure in CMEs does impart a solid body-like character to CMEs \citep{2023debesh1}, which implies that the sheath could be regarded as something anologous to a boundary between the background solar wind and the ``wall'' of the CME front. The high occurrence rate of turbulent spots in sheaths (as compared to the background solar wind) can be studied in this context. The value of our study derives from the size of our sample (125 events), which enables us to draw statistical inferences. Furthermore, we study turbulent spots in $B$, $V$ and $n$, as opposed to many others, which concentrate only on magnetic field turbulence. We note that the data we have used has a relatively low time cadence, and this prevents us from using them for meaningful studies of spectral slopes. Such studies are better done using high cadence data from spacecraft such as the Parker Solar Probe and the Solar Orbiter (e.g., \citep{2026arXiv260106904S}).

\bibliographystyle{plain}

\bibliography{research_proposal_apurva_new}


\end{document}